\documentclass[11pt]{JHEP3}
 \usepackage{amsfonts}
\usepackage{amsmath} 
 \usepackage{amssymb}
 \newfont{\bbbold}{msbm10 scaled \magstep1}

 \def\cF{{\cal F}}

 \def\cM{{\cal M}}

 \newfont{\goth}{eufm10 scaled \magstep1}

 \def\a{\alpha}
 \def\b{\beta}
 \def\c{\gamma}\def\C{\Gamma}
 \def\d{\delta}\def\D{\Delta}
 \def\e{\epsilon}
 
 \def\h{\eta}
 \def\k{\kappa}
 \def\l{\lambda}\def\L{\Lambda}
 \def\m{\mu}

 \def\P{\Pi}

 \def\th{\theta}

 \def\del{\partial}

 \def\ua{\underline{\alpha}}
 \def\ub{\underline{\phantom{\alpha}}\!\!\!\beta}
 \def\uc{\underline{\phantom{\alpha}}\!\!\!\gamma}
 \def\um{\underline{\mu}}
 \def\ud{\underline\delta}
 
 \def\una{\underline a}\def\unA{\underline A}
 \def\unb{\underline b}\def\unB{\underline B}
 \def\unc{\underline c}\def\unC{\underline C}

 \def\unm{\underline m}\def\unM{\underline M}

 \def\nab{\nabla}

 \def\del{\partial}

 \def\be{\begin{equation}}\def\ee{\end{equation}}
 \def\bea{\begin{eqnarray}}\def\eea{\end{eqnarray}}
 \def\ba{\begin{array}}\def\ea{\end{array}}

\title{Kappa-symmetric Deformations of M5-brane Dynamics}

\author{James M. Drummond\\ Department of Mathematics, Trinity College Dublin\\
  E-mail: \email{jmd@maths.tcd.ie} }
\author{Sven F. Kerstan\\ Centre for Theoretical Physics, University of
  Groningen \\ E-mail: \email{S.Kerstan@phys.rug.nl} }

\abstract{We calculate the first
  supersymmetric and kappa-symmetric derivative deformation of the M5-brane
  worldvolume theory in a flat eleven-dimensional background. By applying
  cohomological techniques we obtain a deformation of the standard
  constraint of the superembedding formalism. The first possible deformation
  of the constraint and hence the equations of motion arises at cubic order in
  fields and fourth order in a fundamental length scale $l$. The deformation
  is unique up to this order. In particular this rules out any induced
  Einstein-Hilbert terms on the worldvolume. We explicitly calculate
  corrections to the equations of motion for the tensor gauge supermultiplet.}

\preprint{UG-04-07}

\begin{document}
















\section{Introduction}
In the web of dualities relating ten-dimensional string theories and
eleven-dimensional supergravity, the M5-brane plays a central role. It is
important therefore to understand the structure of the effective field theory
which describes M5-brane dynamics. The multiplet which describes the
linearised worldvolume dynamics of an M5-brane in eleven dimensions is the
on-shell $N=(2,0)$, $D=6$ tensor multiplet \cite{hst83} whose bosonic sector
contains a 
2-form gauge potential with self-dual 3-form field strength as well as five
scalars. The full non-linear equations of motion for the worldvolume theory of
the M5-brane in a general eleven-dimensional supergravity background were
first constructed using the superembedding formalism \cite{hs96,hsw97}. The  
non-linear theory constructed in \cite{hs96,hsw97} is the tensor gauge theory
analogue of supersymmetric Dirac-Born-Infeld theory which is the non-linear
theory describing the worldvolume dynamics of the D-branes of type II string
theory \cite{cvgnw96,aps96,cvgnsw96}. Since one can obtain the IIA theory by
reduction 
of the eleven-dimensional theory on a circle, one can relate the worldvolume
dynamics of the M5-brane to the dynamics of the D4-brane by double dimensional
reduction and indeed it was 
shown in \cite{hsw97} that the non-linear tensor gauge theory correctly
reproduces the Dirac-Born-Infeld equations of motion after such a reduction. 
A covariant action which reproduces the equations of motion was subsequently
constructed in \cite{blnpst97}. 

Just as the Dirac-Born-Infeld theory describing the worldvolume dynamics of
D-branes receives derivative corrections, one expects 
the tensor gauge theory that describes the fluctuations of the M5-brane
to receive derivative corrections. These corrections will involve the
derivative of 
the self-dual 3-form field strength $h_{abc}$, induced curvature
correction terms and higher derivative terms for the fermion fields. We will
show how the form of these derivative corrections can be systematically
constrained by supersymmetry and kappa-symmetry. Our approach is directly
analogous to that of \cite{hklt03,dk04} where it was shown that the
deformations of the standard superembedding constraints can be classified
according to spinorial cohomology \cite{cnt01a,cnt01c}. We will show that the
first possible deformations to the equations of motion are cubic in the fields
and appear at fourth order in a fundamental length parameter $l$.
In particular this implies that there is no induced Einstein-Hilbert type term
in the effective action.

\section{The M5-brane as a superembedding}

We now briefly describe the superembedding formalism \cite{sorokinreview} as
applied to the linearised description of the M5-brane \cite{hs96a}. We
consider the embedding of an $N=(2,0)$, $D=6$ superspace, $\cM$, into a $N=1$,
$D=11$ superspace, $\underline{\cM}$. The coordinates of $\cM$
($\underline{\cM}$) are denoted by $z^M = (x^m, \th^\m)$ ($z^{\unM} =
(x^{\unm}, \th^{\um})$), with Latin indices for bosonic coordinates and Greek
for fermionic. Cotangent frames are related by the vielbein matrices, $E^A =
(E^a , E^{\a}) = dz^M E_{M}{}^{A}$ ($E^{\unA} = (E^{\una} , E^{\ua}) =
dz^{\unM} E_{\unM}{}^{\unA}$). Here the Latin indices $a$ ($\una$) are vector
indices of the group $Spin(1,5)$ ($Spin(1,10)$). The Greek indices $\ua$ are
spinor indices of $Spin(1,10)$ while the indices $\a$ are multi-indices
containing a $Spin(1,5)$ spinor index and a $Spin(5) \cong
Usp(4)$ spinor index.  
The embedding splits target space indices into tangent and normal indices on
the worldvolume, $\unA = (A, A')$. The normal indices $a'$ are $Spin(5)$
vector indices while $\a '$ are again multi-indices containing $Spin(1,5)$ and
$Spin(5)$ spinor indices.  
Later we will introduce the `two-step' notation and explicitly replace a
subscript $\a$ with a subscript pair $\a i$. A subscript $\a '$ is replaced by
a pair $\! \! \! \phantom{a}^{\a}_{i}$. Lower and upper $\a$ indices
distinguish left and right handed Weyl spinors in six dimensions.

The embedding is a map, $f: \cM \longrightarrow
\underline{\cM}$ and it induces the pullback map relating the two frames,

\be
f^* E^{\unA} = E^A E_{A}{}^{\unA}, \hspace{30pt}  
E_{A}{}^{\unA} = E_{A}{}^{M} \del_M z^{\unM} E_{\unM}{}^{\unA}.
\ee

The standard embedding condition $E_{\a}{}^{\una} = 0$ was shown in
\cite{hs96a} to fix the worldvolume multiplet to be the $N=(2,0)$, $D=6$
supersymmetric tensor gauge theory multiplet and it implies the full
non-linear equations of motion \cite{hs96,hsw97} which describe the
fluctuations of the M5-brane in a general eleven-dimensional supergravity
background. Imposing the embedding condition, one can parametrise the
embedding matrix as follows, 

\be
E_{a}{}^{\una} = 
\left(
\begin{array}{l|l}
u_{a}{}^{\una} & \L_{a}{}^{\b '}u_{\b '}{}^{\ua}  \\
\hline
0 & u_{\a}{}^{\ua} + h_{\a}{}^{\b'}u_{\b'}{}^{\ua}
\end{array}
\right).
\ee

Here we have an element of $SO(1,10)$, written $u_{\unb}{}^{\una}$, which is
split according to $u_{\unb}{}^{\una} = (u_{b}{}^{\una},u_{b'}{}^{\una})$. We
denote by $u_{\ub}{}^{\ua}$ the corresponding element of $Spin(1,10)$ which
obeys $(\C^{\unb})_{\ua \ub} u_{\unb}{}^{\una} = u_{\ua}{}^{\uc}
u_{\ub}{}^{\ud} (\C^{\una})_{\uc \ud}$ and is split according to
$u_{\ub}{}^{\ua} = (u_{\b}{}^{\ua} , u_{\b '}{}^{\ua})$.

The relevant information can be derived from the torsion equation which
follows from applying the pullback map to the definition of the target space
torsion, 

\be 
d f^* E^{\unA} = f^* T^{\unA}.
\ee 

We take the target space to be flat $N=1$ $D=11$ superspace with the only
non-zero components of the target space torsion given by

\be
T_{\ua \ub}{}^{\unc} = -i(\C^{\unc})_{\ua \ub}.
\ee

In components the torsion equation reads

\be
\nab_A E_{B}{}^{\unC} - (-1)^{AB} \nab_B E_{A}{}^{\unC} + T_{AB}{}^{C}
E_{C}{}^{\unC} = (-1)^{A(B+\unB)} E_{B}{}^{\unB}E_{A}{}^{\unA} T_{\unA
  \unB}{}^{\unC}. 
\ee

In analysing the torsion equation one encounters the Lie algebra-valued
quantity $X_{A, \una}{}^{\unb} = \nab_A u_{\una}{}^{\unc}
u^{-1}{}_{\unc}{}^{\unb}$. The corresponding quantity with spinorial indices
is related by gamma matrices, 

\be
X_{A,\ua}{}^{\ub} = \nab_A u_{\ua}{}^{\uc} u^{-1}{}_{\uc}{}^{\ub} =
\frac{1}{4} X_{A, \una}{}^{\unb} (\C^{\una}{}_{\unb})_{\ua}{}^{\ub}. 
\ee

We can choose the worldvolume connection so that $X_A$ takes a convenient 
form. The freedom in choosing the connection allows us to set
$\nab_A u_{a}{}^{\unb} u^{-1}{}_{\unb}{}^{b} = \nab_A u_{a'}{}^{\unb}
u^{-1}{}_{\unb}{}^{b'} = 0$. This implies that $\nab_A u_{\a}{}^{\ub}
u^{-1}{}_{\ub}{}^{\b} = \nab_A u_{\a '}{}^{\ub} u^{-1}{}_{\ub}{}^{\b '} = 0$
since these quantities are simply related by gamma matrices. The non-zero
quantities that remain are $X_{A , a}{}^{b'}$ and $X_{A, \a}{}^{\b'} =
\frac{1}{2} X_{A, a}{}^{b'} (\C^a \C_{b'})_{\a}{}^{\b '}$.
As we shall see below $X_{\a, a}{}^{b'}$ is related to the fermion field
$\L_{a}{}^{\a '}$ while $X_{a,b}{}^{c'}$ is the bosonic second fundamental
form.   
  
Analysing the torsion equation level by level in the linearised approximation
we find the supervariations of $h$ (dimension $\frac{1}{2}$), $\L$ (dimension
1) which imply the supervariation of $X$. These are all given below in the
two-step notation,

\be
\nab_{\a i} h_{abc} = -\frac{i}{8} \L_{[a \b i} (\c_{bc]})^{\b}{}_{\a},
\ee

\be
\nab_{\a i} \L_{b \d}{}^{l} = -\frac{1}{2} X_{b,a}{}^{c'} (\c^a)_{\a
  \d} (\c_{c'})_{i}{}^{l} + \nab_b h_{cde} (\c^{cde})_{\a \d} \d_{i}^{l},
\ee

\be
\nab_{\a i} X_{b,a}{}^{c'} = -i (\c^{c'})_{i}{}^{j} \nab_b \L_{a \a j}.
\ee

At dimension $\frac{1}{2}$ we also have the relation of $X_{\a , a}{}^{b'}$
and $\L_{a}{}^{\a '}$ which reads in two-step notation,

\be
X_{\a i , a}{}^{b'} = i \L_{a \a}{}^{j} (\c^{b'})_{ij}.
\ee

We also obtain the linearised equations of motion for the fermions (at
dimension $\frac{1}{2}$) and for the scalars and tensor (at dimension 1),

\be
(\c^a)^{\a \b} \L_{a \a i} = 0,
\ee

\be
\eta^{ab} X_{a,b}{}^{c'} = 0,
\ee

\be
\nab^a h_{abc} = 0.
\ee

The last equation implies a Bianchi identity for $h$ since $h_{abc}$ is
self-dual. In addition we find the constraints $X_{[a,b]}{}^{c'} = \nab_{[a}
\L_{b] \d}{}^{l} = 0$. These equations can be summarised by the statements
that, in the linearised approximation, the fields $h$, $\L$ and $X$ and their
derivatives lie in the following representations of $Spin(1,5) \times
Spin(5)$:   

\begin{align}
\nab_{a_1...a_n}h_{bcd} &\in (n02) \times (00), \label{hfields}\\
\nab_{a_1...a_{n-1}} \L_{a_n \a i} &\in (n01) \times (01), \label{Lfields}\\
\nab_{a_1...a_{n-2}} X_{a_{n-1},a_n}{}^{a'} &\in (n00) \times (10)
\label{Xfields}.
\end{align}

The irreducible representations are given in highest weight notation in the
form $(abc)\times (de)$ with $a,b,c$ Dynkin labels of $Spin(1,5)$ (D3 in the
Cartan classification) and $d,e$ Dynkin labels of $Spin(5)$ (B2 in the Cartan
classification). 

The full non-linear equations which follow from the embedding condition give
the equations of motion, supervariations and components of the worldvolume
torsion to all orders in number of fields as described in
\cite{hs96}. Here we require only the linearised analysis since we use a
perturbative approach as in \cite{dk04} to construct derivative
deformations. Thus we regard the equations of motion as being determined order
by order in terms of the fields of the linearised theory in the
representations given above.   

\section{Derivative corrections}
There is only one way to adapt the analysis of the previous section so that
derivative corrections are included, namely to relax the embedding condition
and allow $E_{\a}{}^{\una}$ to be a function of the fields. This approach was
first used for the membrane in eleven dimensions in \cite{hklt03} and a
similar one (involving deformations of the $\cF$-constraint instead of the
embedding condition) for the D9-brane of IIB in \cite{dk04}.
Deforming $E_{\a}{}^{\una}$ requires the
introduction of an explicit length scale, $l$, since
$E_{\a}{}^{\una}$ has dimension $-\frac{1}{2}$ while the fields $h$, $\L$ and
$X$ have dimensions $0$, $\frac{1}{2}$ and $1$ respectively. 
As in \cite{hklt03} we parametrise the deformation by $\psi_{\a}{}^{a'}$ so
that

\be
E_{\a}{}^{\una} = \psi_{\a}{}^{a'} u_{a'}{}^{\una}.
\ee

The quantity $\psi$ is only defined up to field redefinitions as discussed in
\cite{hklt03} . The basic fields are the embedding coordinates $z^{\unM}$ and
the effect of the redefinition $z^{\unM} \longrightarrow z^{\unM} + (\d
z)^{\unM}$ will define the ambiguity in $\psi$. The field redefinition is
equivalent to a target space diffeomorphism and under such a transformation,
given by a vector field $v$, the target space frame transforms as

\be
\d_v E^{\unA} = (d i_v + i_v d) E^{\unA} = d v^{\unA} + E^{\unC} v^{\unB}
T_{\unB \unC}{}^{\unA}.
\ee

Applying the pullback map gives the transformation of the embedding matrix,

\be
\d_v E_{A}{}^{\unA} = \nab_A v^{\unA} + E_{A}{}^{\unC} v^{\unB} T_{\unB
  \unC}{}^{\una}. 
\ee

Setting $A =\a$ and $\unA = \una$ we find the transformation of $\psi$,

\be
\d_v \psi_{\a}{}^{c'} = \nab_\a \hat{v}^{c'} + \hat{v}^b X_{\a, b}{}^{c'} -i
\hat{v}^{\d '} (\C^{c'})_{\d' \a} -i h_{\a}{}^{\e '} \hat{v}^\d (\C^{c'})_{\d
  \e '}, 
\ee

where $\hat{v}^{\una} = v^{\unb} u_{\unb}{}^{\una}$ and $\hat{v}^{\ua} =
v^{\ub} u_{\ub}{}^{\ua}$. 

Choosing $\hat{v}^a = \hat{v}^{\a} = 0$ gives

\be
\d_v \psi_{\a}{}^{c'} = \nab_\a \hat{v}^{c'} -i \hat{v}^{\d '} (\C^{c'})_{\d '
  \a}. 
\ee

The second term allows us to remove the gamma-trace part of $\psi$ so that we
only need to look for deformations in the $(001) \times (11)$ representation
of $Spin(1,5) \times Spin(5)$. The first term implies that those
$\psi_{\a}{}^{c'}$ which are given by $\nab_{\a} V^{c'}$ (projected onto the
$(001) \times (11)$ representation) for some vector $V^{c'}$ are trivial
deformations which can be removed by field redefinitions.

Hence we find the equivalence,

\be
\psi_{\a}{}^{c'} \cong \psi_{\a}{}^{c'} + \nab_{\a} V^{c'}.
\ee

As well as the field redefinitions which describe the above ambiguity in
$\psi$ there are constraints which $\psi$ must satisfy. We find these by
examining the dimension zero part of the torsion equation in the presence of
non-zero $\psi$ which reads

\be
\nab_{\a i} (\psi_{\b j}{}^{c'} u_{c'}{}^{\unc}) + (\b j \leftrightarrow
\a i) + T_{\a i \b j}{}^{c} u_{c}{}^{\unc} + T_{\a i \b j}{}^{\c k} \psi_{\c
  k}{}^{d'}u_{d'}{}^{\unc} = E_{\a i}{}^{\ua} E_{\b j}{}^{\ub} T_{\ua
  \ub}{}^{\unc}.  
\ee 

Applying $u_{\unc}{}^{c}$ gives the dimension zero part of the worldvolume
torsion.
Applying $u_{\unc}{}^{c'}$, we find the constraints which $\psi$ must
satisfy for the deformation to be consistent,

\be
\nab_{\a i} \psi_{\b j}{}^{c'} + (\b j \leftrightarrow \a i) = -i
(\c^{c'})_{il} h_{\b j \a}{}^{l} + (\b j \leftrightarrow \a i).
\ee

We split $h_{\a i \b}{}^{j}$ into its irreducible representations,

\begin{align}
h_{\a i \b}{}^{j} = &\d_{i}^{j} [h^a(\c_a)_{\a \b} + h^{abc}(\c_{abc})_{\a
    \b}] \notag \\ 
+&(\c^{a'})_{i}{}^{j} [h_{a'}^{a} (\c_a)_{\a \b} + h_{a'}^{abc} (\c_{abc})_{\a
    \b}] \notag \\ 
+ &(\c^{a'b'})_{i}{}^{j} [h_{a'b'}^{a} (\c_a)_{\a \b} +  h_{a'b'}^{abc}
    (\c_{abc})_{\a \b}]. 
\end{align}

In the case $\psi=0$ we then find that only $h^{abc}$ is non-zero. For general
$\psi$ we find the constraints

\begin{align}
Y_{a}^{a';c'} &= (\c_a)^{\a \b} (\c^{a'})^{ij} \nab_{\a i} \psi_{\b j}{}^{c'}
= 0, \label{Yconstraint} \\
Z_{abc}^{a'b';c'} &= (\c_{abc})^{\a \b} (\c^{a'b'})^{ij} \nab_{\a i} \psi_{\b
  j}{}^{c'} = 0, \label{Zconstraint}
\end{align}

where in the first line it is to be understood that one keeps only the $(100)
\times (20)$ representation of $Spin(1,5) \times Spin(5)$ and in the second
line only the $(002) \times (12)$ representation.

At lowest order in the deformation and lowest order in the number of fields,
the algebra of spinorial derivatives is the standard flat superspace algebra,

\be
[\nab_{\a i}, \nab_{\b j}] = i\eta_{ij}(\c^a)_{\a \b}\nab_a, \label{derivalg} 
\ee

so that any $\psi_{\a i}{}^{c'}$ of the form $\nab_{\a i} V^{c'}$ satisfies
the constraints (\ref{Yconstraint},\ref{Zconstraint}) automatically up to
higher orders. In the above formula $\eta_{ij}$ is the antisymmetric $Spin(5)
\cong Usp(4)$ invariant antisymmetric tensor.

Thus we can consider the sequence of representations of $Spin(1,5) \times
Spin(5)$,

\be
(000) \times (10) \phantom{a}^{\D} \!\!\!\!\!\!\!\! \longrightarrow
(001) \times (11) \phantom{a}^{\D} \!\!\!\!\!\!\!\! \longrightarrow
((100) \times (20)) \oplus ((002)\times (12)).
\label{5brseq}
\ee

The operation $\D$ is given by applying a spinorial derivative and projecting
onto the target representation. The algebra (\ref{derivalg}) implies $\D$ is
nilpotent, $\D^2=0$. The analysis above can be summarised by saying that the
genuine deformations of $\psi$ away from zero are $\D$-closed (they
satisfy the constraints) and equivalent if they differ by $\D$-exact terms
(field redefinitions). Hence we are looking for $\psi$ in the cohomology of
$\D$, 

\be
\psi \in H = \frac{{\rm Ker} \D}{{\rm Im} \D} .
\ee   

We now explicitly calculate the cohomology $H$, working order by order in the
number of fields. We must construct $\psi$ in the representation $(001) \times
(11)$ from the fields and their derivatives constrained by the linearised
analysis (\ref{hfields}, \ref{Lfields}, \ref{Xfields}). Also we must check
possible field redefinitions in the representation $(000) \times (10)$ and
constraints in the representations $(100) \times (20)$ ($Y$-constraints) and
$(002) \times (12)$ ($Z$-constraints). It is obvious that
there can be nothing for $\psi$ linear in the fields since there is nothing in
the correct representation. In fact this is already enough to rule out induced
Einstein-Hilbert terms in the effective action.

At quadratic order in fields we find

\be
\psi_{\a i}{}^{a'} = l^{2n-1} \nab_{a_1}...\nab_{a_{n-1}} \L_{a_n \a i}
\nab^{a_1}...\nab^{a_{n-2}} X^{a_{n-1},a_n a'}, \text{ with } n \geq 2.
\ee

There are no field redefinitions $V^{a'}$ quadratic in the fields but there
are constraints of the $Y$-type,

\be
Y_{a}^{a';b'} = l^{2n-1} \nab_a \nab_{a_1}...\nab_{a_n-2}
X_{a_{n-1},a_n}{}^{a'} \nab^{a_1}...\nab^{a_{n-2}} X^{a_{n-1},a_n b'}
\text{ with } n \geq 2.
\ee

It is then simple to see that for each $n$ applying a spinorial derivative and
projecting onto the $Y$-representation gives a non-zero answer and hence none
of the quadratic $\psi$ are $\D$-closed. Thus the cohomology quadratic in
fields is trivial. 

Moving on to cubic order in fields we find the first possibilities at order
$l^2$. There are three linearly independent deformations, $\psi$, two possible
field redefinitions, $V$, and two constraints, one each of $Y$-type and
$Z$-type. The explicit formulae for these are given in Appendix A. We find that
two linearly independent combinations of the deformations can be removed by
field redefinitions and the remaining combination gives a non-zero
contribution to the constraints. Thus the cubic cohomology at order $l^2$ is
trivial.  

There is nothing one can write down for $\psi$ at order $l^3$ which is cubic
in fields and so the next
order to check is $l^4$. Here we find 18 linearly independent deformations,
$\psi$. There are 8 possible field redefinitions, $V$ and 16 constraints, 8
each of $Y$-type and $Z$-type. The details of these are given in Appendix
B. At this order we find that 8 linearly independent combinations of the
deformations can be removed by field redefinitions. Of the remaining 10
combinations only 2 are closed in the $Y$-sense. Finally, of these 2 only one
is also closed in the $Z$-sense. It can be written,

\begin{align}
\psi_{\a i}{}^{a'} = l^4[&4i\L_{a \a}{}^{j} \nab_b X_{c}{}^{a b'} X^{b,c
  a'}(\c_{b'})_{ij} \notag \\
+& \nab_a \nab_b \L_{c \a}{}^{j} \L^{a}{}_{\c}{}^{k}
  \L^{b}{}_{\d}{}^{l} (\c^{a'b'})_{kl} (\c_{b'})_{ij} (\c^c)^{\c \d} \notag \\ 
+&24i
  \L_{a \b i} X_{b,c}{}^{a'} \nab^a \nab^b h^{cde} (\c_{de})_{\a}{}^{\b}] .
\label{deform}
\end{align} 

The above formula is the first supersymmetric deformation of the embedding
condition for the M5-brane in a flat eleven-dimensional background. The fact
that the deformation is cubic in fields implies that the resulting corrections
to the equations of motion are also cubic in fields. In particular this means,
in terms of effective actions, that the pure curvature corrections will be
quartic in the second fundamental form, $X$, and hence quadratic in the
Riemann curvature, $R \sim X^2$. We will now show how the equations of motion
of the deformed theory can be calculated from the above formula.

\section{Derivative corrections to the equations of motion}

One can derive the deformed equations of motion by analysing the torsion
equation level by level in dimension in the presence of a non-zero $\psi$. As
we have seen at dimension zero this implies constraints that $\psi$ must
satisfy. It also determines the remaining representations present in $h_{\a i
  \b}{}^{j}$,
\begin{align}
h_a &= 0, \\
h_{a}^{a'} &= \frac{i}{16} \eta^{ij} (\c_a)^{\a \b} \nab_{\a i} \psi_{\b
  j}{}^{a'}, \\
h_{a}^{a'b'} &= \frac{i}{32} (\c^{[a'})^{ij} (\c_a)^{\a \b} \nab_{\a i}
  \psi_{\b j}{}^{b']}, \\
h_{abc}^{a'} &= \frac{i}{2.4^3 .6}(\c^{a'b'})^{ij} (\c_{abc})^{\a \b} \nab_{\a
  i} \psi_{\b j b'}, \\
h_{abc}^{a'b'} &= \frac{i}{2^2. 4^2. 6^2} \e^{a'b'c'd'e'}(\c_{abc})^{\a \b}
(\c_{c'd'})^{ij} \nab_{\a i} \psi_{\b j e'}.
\end{align} 

At dimension $\frac{1}{2}$ we find from the $\!\!\!\phantom{a}_{\a i
  b}{}^{\unc}$ part of the torsion equation, contracted with
  $(u^{-1})_{\unc}{}^{c'}$, that the relation between $X_{\a i,
  b}{}^{c'}$ and $\L$ is modified,
\be
X_{\a i, b}{}^{c'} = i \L_{b \a}{}^{k} (\c^{c'})_{ik} + \nab_b \psi_{\a
  i}{}^{c'}. 
\ee
Note that the first term also contains a correction term, the gamma trace of
  $\L$, which is only zero at lowest order (the linearised fermionic equation
  of motion). We write
\be
\L_{a \a i} = (\c_a)_{\a \b} \P_{i}^{\b} + \hat{\L}_{a \a i}, 
\ee
where $\hat{\L}_{a \a i}$ is gamma-traceless and $\P$ is the correction to the
fermionic equation of motion. Employing these relations we also find from the
  $\!\!\!\phantom{a}_{\a i \b j}{}^{\uc}$ part of the torsion equation, 
\begin{align}
[&-\frac{i}{2} (\c_b)_{\a \e} \P^{\e k} (\c^{c'})_{ik} (\c^b)_{\b \d}
  (\c_{c'})_{j}{}^{l} - \frac{1}{2} \nab_b \psi_{\a i}{}^{c'} (\c^b)_{\b \d}
  (\c_{c'})_{j}{}^{l} \notag \\
&+ (\nab_{\a i} h_{\b j \d}{}^{l})^{(3,1)}] +{\a i
  \leftrightarrow \b j} -i(\c^c)_{\a \b} \eta_{ij} (\c_c)_{\d \e} \P^{\e l} =0.
\end{align}
The superscript $(3,1)$ means the part which is cubic in fields and first
order in the deformation parameter, which for the term we have calculated is
$l^4$. Contracting with $(\c_d)^{\a \b} \eta^{ij}$ we find
\begin{align}
& -  (\nab_{\a}{}^{l} h_{abc})^{(3,1)} (\c_d
\c^{abc})^{\a}{}_{\d} +(\c^{a'})^{il} (\c_d \c_a)^{\a}{}_{\d} \nab_{\a i}
h_{a'}^{a} +(\c^{a'})^{il} (\c_d \c_{abc})^{\a}{}_{\d} \nab_{\a
  i}h_{a'}^{abc} \notag \\
&+(\c^{a'b'})^{il} (\c_d \c_a)^{\a}{}_{\d} \nab_{\a i} h_{a'b'}^{a} +
(\c^{a'b'})^{il} (\c_d \c_{abc})^{\a}{}_{\d} \nab_{\a i} h_{a'b'}^{abc} 
+18i(\c_d)_{\d \e} \P^{\e l}=0.
\end{align}
Taking the gamma-trace by contracting with $(\c^d)^{\d \h}$ gives
\be
27i\P^{\h l} = - (\c_a)^{\a \h} (\c^{a'})^{il} \nab_{\a i} h_{a'}^{a} -
(\c^{a'b'})^{il} (\c_a)^{\a \h} \nab_{\a i} h_{a'b'}^{a},
\ee
which is the correction to the fermionic equation of motion.

At dimension one we find from the $\!\!\!\phantom{a}_{\a i b}{}^{\uc}$ part of
the torsion equation, 
\be
(\nab_{\a i} \L_{d \c}{}^{k})^{(3,1)} = (\nab_d h_{\a i \c}{}^{k})^{(3,1)} +
(X_{d, \a i \c}{}^{k})^{(3,1)}.
\ee
Contracting with $\d^{i}_{k} (\c_{ab} \c^d)^{\a \c}$ gives the
  correction to the 
tensor field equation,
\be
(\nab^c h_{abc})^{(3,1)} = \frac{1}{32} \nab_{\a i} \P^{\d i}
(\c_{ab})_{\d}{}^{\a}. 
\ee
Contracting with  $(\c^{a'})_{k}{}^{i} (\c^d)^{\a \c}$ gives the correction to
the scalar equation of motion,
\be
(\eta^{bc} X_{b,c}{}^{a'})^{(3,1)} = \frac{3}{4} \nab_{\a i} \P^{\a k}
  (\c^{a'})_{k}{}^{i} + 2 \nab_b h^{b a'}.
\ee

Thus all the corrections to the equations of motion are specified once $\psi$
is known. As an example we give here the pure tensor field corrections to the
tensor field equation, 
\be
\nab^a h_{abc} =  120 l^4  (2\nab_a \nab_d \nab_e h_{bcf} \nab^a h_{gh}{}^{f}
\nab^d h^{egh} + 5 \nab_a h_{de}{}^{f} \nab_g \nab_h h_{bcf} \nab^a \nab^g
h^{hde}).
\ee

We should stress that although we have given purely bosonic terms for
simplicity, all corrections to the equations of motion are specified by this
method, including all fermion terms. 

\section{Relation to deformations of $N=(2,0)$, $D=6$ tensor gauge theory}
There is an alternative way to derive deformations of
the dynamics of the M5-brane by applying the spinorial
cohomology techniques directly, without the superembedding
\cite{cnt01c}. In this section we sketch how the first deformation,
corresponding to the leading contributions in the non-linear
field equations (without derivative corrections)
is derived this way. In principle our results of the previous sections could be
derived from this approach, but we hope to illustrate with this
short recap that the simpler and more powerful approach is to deform
the superembedding, because we get the non-linear (non-derivative) corrections "for free" and, additionally, the superembedding approach we presented keeps $\k$-symmetry and reparametrisation invariance manifest. 

To derive the deformations of the previous section one could directly deform the superspace constraints for the $N=(2,0)$, $D=6$ tensor multiplet \cite{hst83}. This
multiplet can be described by a closed 3-form, $H$, on $N=(2,0)$ , $D=6$
superspace whose non-zero components are
\begin{align}
H_{\a i \b j c} &= -i (\c_c)_{\a \b} W_{ij}, \\
H_{\a i b c} &= (\c_{bc})_{\a}{}^{\b} \l_{\b i}, 
\end{align}
as well as $H_{abc}$ which is self-dual. The fields are constrained to satisfy
their equations of motion,
\be
\del_a \del^a W_{ij} = (\c^a)^{\a \b}\del_a \l_{\b k} = \del^c H_{abc} = 0.
\ee

The constraint $H_{\a i \b j \c k} = 0$ and the closure condition
$dH=0$ imply the fields of the tensor multiplet satisfy their equations of
motion. 
The worldvolume theory of the M5-brane is described by a deformation of the
constraints for the $N=(2,0)$ tensor multiplet. If one ignores derivative
corrections then the full non-linear deformation of the constraint $H_{\a i \b
  j \c k} = 0$ which defines the non-linear tensor gauge theory of the
M5-brane must be implied by the embedding condition of the superembedding
formalism since the constraint $E_{\a}{}^{\una}=0$ defines the full non-linear
theory \cite{hs96}. The analogous relation between the $\cF$-constraint of the
superembedding formalism and the non-linear deformation of the $D=10$ Abelian
Yang-Mills constraints which defines Born-Infeld theory was developed in
\cite{kerstan02}. There the relation is much more direct since the two
constraints take a similar form; they are both constraints on a superspace
2-form. Here the constraints are on objects of different dimension and carrying
different representations of $Spin(1,5) \times Spin(5)$, $\psi_{\a i}{}^{a'}$
which is of dimension $-\frac{1}{2}$ and $H_{\a i \b j \c k}$ which (if we
stick to the dimensions we have used throughout) is of dimension
$-\frac{3}{2}$. This situation is similar to the different descriptions of
eleven-dimensional supergravity where one can apply superspace constraints
to the torsion 2-form or, in the 4-form formulation, to the superspace
4-form. 

In terms of cohomology the deformation of the tensor multiplet constraints was
studied in \cite{cnt01c}. The relevant representations of $Spin(1,5) \times
Spin(5)$ in $H_{\a i \b j \c k}$ which remain after conventional constraints
are $(003) \times (03)$ and $(101) \times (11)$. The relevant field
redefinitions, which are redefinitions of the purely spinorial part of the
2-form potential lie in the representations $(002) \times (02)$ and $(100)
\times (10)$. Finally the constraints which follow from the Bianchi identity
lie in the representations $(004) \times (04)$, $(102) \times (12)$ and $(200)
\times (20)$. We thus have the complex,

\be
\begin{array}{lllll}
(002) \times (02) & \longrightarrow & (003) \times (03) & \longrightarrow &
  (004) \times (04) \\
&\searrow & &\searrow & \\
(100) \times (10) & \longrightarrow & (101) \times (11) & \longrightarrow &
  (102) \times (12) \\
&  & & \searrow & \\
& & & & (200) \times (20).
\end{array}
\ee
The arrows represent operators given by applying a spinorial derivative and
projecting onto the target representation.
The non-trivial deformations of the constraints lie in the cohomology of the
combined operation defined on the reducible sum of the irreps in each column.

The fields of the tensor multiplet consist of the scalars, $W_{ij}$ in the
$(000) \times (10)$ representation, fermions $\l_{\a i}$ in the $(001)
\times (01)$ representation and the self-dual field strength $H_{abc}$ in the
$(002) \times (00)$ representation. In our conventions these have dimensions
$-1$, $-\frac{1}{2}$ and $0$ respectively.  
By dimensional analysis and inspection of the possible representations we can
see that the leading deformation is cubic in fields and of the form $\l W H$
or $\l \l \l$. This induces terms cubic in the field strength $H_{abc}$
(but with no derivatives) in the tensor field equation as well as accompanying
scalar and fermion terms. Such a deformation corresponds to the leading
non-derivative correction, the analogue of the leading Born-Infeld correction
for the tensor gauge theory.   
In the manifestly kappa-symmetric approach described in the previous sections
these corrections are completely taken into account by the embedding
condition. The deformations of the embedding condition, which we have
discussed correspond to deformations of the tensor multiplet constraints which
include derivatives and an explicit dimensionful parameter, $l$. It is
interesting to note that the kappa-symmetric formulation of the problem is
much more tractable than the standard cohomology approach for deforming and
solving the Bianchi identity for the tensor multiplet.

\section{Conclusions}
We have shown how spinorial cohomology can be used to derive the leading
derivative and curvature corrections to the equations of motion for the tensor
supermultiplet which describes the worldvolume dynamics of the M5-brane. The
method used is manifestly supersymmetric and kappa-symmetric as these
symmetries are implemented geometrically by using superspaces for the
eleven-dimensional background and six-dimensional worldvolume. 

The corrections we have calculated are fixed by the deformation (\ref{deform})
of the embedding condition of the superembedding formalism. The deformation
was derived working first order by order in number of fields, where it was
shown that there are no deformations at first or second orders, and then order
by order in the scale $l$. The first cubic terms appear at order $l^4$ which
is consistent with previous results on the membrane \cite{hklt03} and open
string theories and D-branes
\cite{at88,bbg99,wyllard00,cdre02,dhhk03,dre03,dk04} and indeed one could
perform a double dimensional reduction of the M5-brane results to obtain those
for the D4-brane of IIA. 

One can continue the calculation presented here to higher order in fields and
higher orders in $l$. It is interesting to note that in the case of the
membrane there are no gauge field degrees of freedom in the worldvolume
multiplet and therefore there is no dimension zero field strength which could
appear in the deformation. This means that there are a finite number of
possible terms for the deformation at any given order in $l$. Thus it is
possible to calculate the complete deformation at a given order in $l$
although the calculation of \cite{hklt03} restricts to cubic terms to make the
computation more tractable. In the case of the M5-brane and D-branes which
contain a gauge field in their worldvolume multiplet this is no longer the
case and one is in principle restricted to performing the analysis order by
order in fields as well as order by order in $l$. However it may be the case
that one need only consider a finite number of combinations involving the
dimension zero field strength and deduce results for all orders in fields at a
given order in $l$, just as one can for the undeformed superembedding.

The calculation here has been performed in a flat background but one can
easily generalise to general on-shell backgrounds of eleven-dimensional
supergravity. In particular one can address the issue of kinetic terms for the
pullbacks of the background supergravity gauge fields in this approach while
preserving all of the supersymmetry and kappa-symmetry. The corresponding
terms for NS-NS and R-R fields for D-branes in type II theories can also be
derived either by dimensional reduction or directly using the same
method. Alternatively one can obtain correction terms by calculating string
amplitudes directly as in
\cite{bbg99,clr01,fotopoulos01,aagg02,wijnholt03,epple04}.  Such terms are
relevant for the quantum polarisation discussed in \cite{cms04}.

\appendix

\section{Cubic cohomology at order $l^2$}

\subsection*{Deformations}
The deformations are given by $\psi_{\a i}{}^{a'}$ in the
  $(001)\times (11)$ irrep of $Spin(1,5) \times Spin(5)$. We have:
\begin{align}
\psi_1 &= \L_{a \b i} X^{a,b a'} h_{bcd} (\c^{cd})_{\a}{}^{\b}, \\
\psi_2 &= \L_{a \b i} X_{b,c}{}^{a'} h^{abd}
(\c^{c}{}_{d})_{\a}{}^{\b}, \\
\psi_3 &= \L_{a \b}{}^{j} \L_{b \c}{}^{k} \L_{c \d i} (\c^c)^{\b \c}
(\c^{ab})_{\a}{}^{\d} (\c^{a'})_{jk}.
\end{align}

\subsection*{Field Redefinitions}
The field redefinitions are given by $v^{a'}$ in the $(000)\times
(10)$ irrep of $Spin(1,5)\times Spin(5)$,
\begin{align}
V_1 &= X_{a,b}{}^{a'} h^{acd} h^{b}{}_{cd} , \\
V_2 &= \L_{a \a}{}^{i} \L_{b \b}{}^{j} h^{abc} (\c_c)^{\a \b} (\c^{a'})_{ij}.
\end{align}

\subsection*{Constraints}
There are two irreps which contribute to the constraints. They are
given by $Y_{a}{}^{a';b'}$ which is the $(100)\times (20)$ irrep of
$Spin(1,5) \times Spin(5)$ and $Z_{abc}{}^{a';b'c'}$ which is the
$(002) \times (12)$ irrep. At this order we have one of each:
\be
Y_{a}{}^{a';b'} = \L_{a \a i} \L_{b \b j} X^{b,c a'} (\c^{b'})^{ij}
(\c_c)^{\a \b},
\ee 
\be 
Z_{abc}{}^{a';b'c'} = \L_{[a \a i} \L_{|b| \b j} X^{b}{}_{c}{}^{a'}
  (\c^{b'c'})^{ij} (\c_{d]})^{\a \b} + \text{ dual }. 
\ee

\subsection*{Spinorial relations}
We apply a spinorial derivative followed by projection (which denote
by $\D$). We find:
\be
\D V_1 = -\frac{i}{12} (\psi_1 + 2 \psi_2), 
\hspace{30pt}
\D V_2 = -2\psi_2 +\frac{i}{24}(\psi_3),
\ee
and
\be
\D \psi_1 = -i Z -\frac{i}{2} Y, 
\hspace{30pt} 
\D \psi_2 = \frac{i}{2} Z + \frac{i}{4} Y,
\hspace{30pt}
\D \psi_3  = 24 Z + 12 Y.
\ee
We see that we can use the two field redefinitions to eliminate two of the
deformations and that the remaining one will have a non-zero contribution to
the constraints and so is not closed. Therefore the cohomology at this order
is trivial.

\section{Cubic cohomology at $l^4$}
\subsection*{Deformations}
We have
\begin{align}
\psi_1 &= \L_{a \a}{}^{j} \nab_b X_{c}{}^{a b'} X^{b,c
  a'}(\c_{b'})_{ij}, \\
\psi_2 &= \L_{a \a}{}^{j} \nab_b X_{c}{}^{a a'} X^{b,c b'}
  (\c_{b'})_{ij}, \\
\psi_3 &= \L_{a \b}{}^{j} \nab^a X_{b,c}{}^{b'} X^{b,d a'}
  (\c_{b'})_{ij} (\c^{c}{}_{d})_{\a}{}^{\b}, \\
\psi_4 &= \L_{a \b}{}^{j} \nab^a X_{b,c}{}^{a'} X^{b,d b'}
  (\c_{b'})_{ij} (\c^{c}{}_{d})_{\a}{}^{\b}, \\
\psi_5 &= \nab_a \L_{b \a}{}^{j} X^{a,c a'}
  X^{b}{}_{c}{}^{b'}(\c_{b'})_{ij}, \\
\psi_6 &= \nab_a \L_{b \b}{}^{j} X^{a,c a'} X^{b,d b'} (\c_{b'})_{ij}
  (\c_{cd})_{\a}{}^{\b}, \\
\psi_7 &= \L_{a \b i} \nab^a X_{b,c}{}^{a'} \nab^b h^{cde}
  (\c_{de})_{\a}{}^{\b}, \\
\psi_8 &= \L_{a \b i} \nab_b X_{c,d}{}^{a'} \nab^b h^{cae}
  (\c^{d}{}_{e})_{\a}{}^{\b} , \\
\psi_9 &= \nab_a \L_{b \b i} \nab^a X^{b,c a'} h_{cde}
  (\c^{de})_{\a}{}^{\b}, \\
\psi_{10} &= \nab_a \L_{b \b i} \nab^a X_{c,d}{}^{a'} h^{bce}
  (\c^{d}{}_{e})_{\a}{}^{\b}, \\
\psi_{11} &= \nab_a \L_{b \b i} X^{a,c a'} \nab^b h_{cde}
  (\c^{de})_{\a}{}^{\b}, \\
\psi_{12} &= \nab_a \L_{b \b i} X_{c,d}{}^{a'} \nab^a h^{bce}
  (\c^{d}{}_{e})_{\a}{}^{\b}, \\
\psi_{13} &= \L_{a \b i} X_{b,c}{}^{a'} \nab^a \nab^b h^{cde}
  (\c_{de})_{\a}{}^{\b}, \\
\psi_{14} &= \nab_a \nab_b \L_{c \a}{}^{j} \L^{a}{}_{\c}{}^{k}
  \L^{b}{}_{\d}{}^{l} (\c^{a'b'})_{kl} (\c_{b'})_{ij} (\c^c)^{\c \d}, \\
\psi_{15} &= \nab_a \L_{b \b}{}^{j} \nab^a \L^{c}{}_{\c}{}^{k} \L_{d \d
  i} (\c^{a'})_{jk} (\c^d)^{\b \c} (\c^{b}{}_{c})_{\a}{}^{\d}, \\
\psi_{16} &= \nab_a \L_{b \b}{}^{j} \nab^a \L^{c}{}_{\c}{}^{k}
  \L^{b}{}_{\d i} (\c^{a'})_{jk} (\c^d)^{\b \c} (\c_{cd})_{\a}{}^{\d},
  \\
\psi_{17} &= \nab_a \L_{b \b}^{j} \nab^a \L^{b}{}_{\c}{}^{k} \L_{c
  \a}{}^{l} (\c^{a'b'})_{jk} (\c_{b'})_{il} (\c^c)^{\b \c}, \\
\psi_{18} &= \nab_a \L_{b \b}{}^{j} \nab^a \L^{c}{}_{\c}{}^{k} \L_{c
  \d}{}^{l} (\c^{a'b'})_{jk} (\c_{b'})_{il} (\c_d)^{\b \c}
  (\c^{bd})_{\a}{}^{\d}. 
\end{align}

\subsection*{Field Redefinitions}
There are eight of these:
\begin{align}
V_1 &= X_{a,b}{}^{a'} X^{b,c b'} X_{c,}{}^{a}{}_{b'}, \\
V_2 &= X_{a,b}{}^{a'} \nab_c h^{ade} \nab^c h^{b}{}_{de}, \\
V_3 &= \nab_a X_{b,c}{}^{a'} h^{ade} \nab^b h^{c}{}_{de}, \\
V_4 &= \L_{a \a}{}^{i} \L_{b \b}{}^{j} \nab^a X^{b,c b'} (\c^{a'}{}_{b'})_{ij}
(\c_c)^{\a \b}, \\
V_5 &= \nab_a \L_{b \a}{}^{i} \L^{a}{}_{\b i} X^{b,c a'} (\c_c)^{\a \b}, \\
V_6 &= \nab_a \L_{b \a}{}^{i} \L^{a}{}_{\b}{}^{j} X^{b,c b'} (\c_c)^{\a \b}
(\c^{a'}{}_{b'})_{ij}, \\
V_7 &= \nab_a \L_{b \a}{}^{i} \L_{c \b}{}^{j} \nab^a h^{bcd} (\c_d)^{\a \b}
(\c^{a'})_{ij}, \\
V_8 &= \nab_a \L_{b \b}{}^{j} \nab^a \L_{c \c}{}^{k} h^{bcd} (\c_d)^{\b \c}
(\c^{a'})_{jk}.
\end{align}

\subsection*{Constraints}
There are sixteen constraints, eight in each representation. For the
representation $(100) \times (20)$ we have,
\begin{align}
Y_1 &= \nab_b X_{c,d}{}^{a'} X^{b,e b'} \nab^c h^{d}{}_{ea}, \\
Y_2 &= \nab_b \L_{c \a i} \L_{d \b j} \nab^b X^{c,d a'} (\c^{b'})^{ij}
(\c_a)^{\a \b}, \\
Y_3 &= \nab_a \L_{b \a i} \L_{c \b j} \nab^b X^{c,d a'} (\c^{b'})^{ij}
(\c_d)^{\a \b}, \\
Y_4 &= \nab_b \L_{c \a i} \L_{a \b j} \nab^b X^{c,d a'} (\c^{b'})^{ij}
(\c_d)^{\a \b}, \\
Y_5 &= \nab_b \L_{c \a i} \L^{b}{}_{\b j} \nab^c X_{a,d}{}^{a'} (\c^{b'})^{ij}
(\c^d)^{\a \b}, \\
Y_6 &= \nab_b \nab_c \L_{d \a i} \L^{b}{}_{\b j} X^{c,d a'} (\c^{b'})^{ij}
(\c_a)^{\a \b}, \\
Y_7 &= \nab_a \nab_b \L_{c \a i} \L^{b}{}_{\b j} X^{c,d a'} (\c^{b'})^{ij}
(\c_d)^{\a \b}, \\
Y_8 &= \nab_a \L_{b \a i} \nab^b \L^{c}{}_{\b j} X_{c,d}{}^{a'} (\c^{b'})^{ij}
(\c^d)^{\a \b}. 
\end{align}
For the representation $(002) \times (12)$ we have,
\begin{align}
Z_1 &= X_{[a,}{}^{da'} X_{b}{}^{eb'} \nab_{c]}X_{d,e}{}^{c'} + \text{ dual, }
\\
Z_2 &= \nab_d \L_{e \a i} \L_{[a \b j} \nab_b X^{d,e c'} (\c^{a'b'})^{ij}
  (\c_{c]})^{\a \b} + \text{ dual, } \\
Z_3 &= \nab_d \L_{[a \a i} \L_{|e| \b j} \nab_b X^{d,e c'} (\c^{a' b'})^{ij}
  (\c_{c]})^{\a \b} + \text{ dual, } \\
Z_4 &= \nab_d \L_{[a \a i} \L_{b \b j} \nab_{c]} X^{d,e c'} (\c^{a'b'})^{ij}
(\c_e)^{\a \b} + \text{ dual, } \\
Z_5 &= \nab_d \L_{e \a i} \L_{f \b j} \nab^d X^{e,f c'} (\c^{a'b'})^{ij}
(\c_{abc})^{\a \b} \\
Z_6 &= \nab_d \nab_e \L_{[a \a i} \L^{d}{}_{\b j} X_{b,}{}^{e c'}
  (\c^{a'b'})^{ij} (\c_{c]})^{\a \b} + \text{ dual, } \\
Z_7 &= \nab_d \nab_e \L_{f \a i} \L^{d}{}_{\b j} X^{e,f c'} (\c^{a'b'})^{ij}
(\c_{abc})^{\a \b}, \\
Z_8 &= \nab_d \L_{e \a i} \nab^d \L_{[a \b j} X^{e,}{}_{b}{}^{c'}
  (\c^{a'b'})^{ij} (\c_{c]})^{\a \b} + \text{ dual.}
\end{align}

\subsection*{Spinorial relations}
Applying a spinorial derivative and a projection to $V_1,...,V_8$ we find
\begin{align}
\D V_1 &= 2i \psi_5, \\
\D V_2 &= \frac{i}{12}[\psi_{11} + 2\psi_{12}], \\
\D V_3 &= \frac{i}{24}[\psi_7 - 2\psi_8 + \psi_9 + 2\psi_{10}], \\
\D V_4 &= -2 \psi_1 + 2 \psi_3 + \psi_2 - \psi_4 + 12\psi_7 + i\psi_{14}, \\
\D V_5 &= \frac{1}{2} \psi_1 + \frac{1}{2} \psi_3 + 6 \psi_{13} - \frac{1}{2}
\psi_5 + \frac{1}{2} \psi_6 - 6 \psi_{11}, \\
\D V_6 &= - \psi_2 - \psi_4 + \frac{1}{2} \psi_1 + \frac{1}{2} \psi_3 + 6
\psi_{13} - \frac{1}{2} \psi_5 - \frac{3}{2} \psi_6 + 6\psi_{11} -
\frac{3i}{4}\psi_{16} - \frac{i}{4} \psi_{18}, \\
\D V_7 &= \psi_8 - \psi_{12} - \frac{i}{48} \psi_{15} - \frac{i}{96}\psi_{16}
- \frac{i}{48} \psi_{17} + \frac{i}{96} \psi_{18}, \\
\D V_8 &= -2 \psi_{10} + \frac{i}{12} \psi_{16} + \frac{i}{24} \psi_{15}.
\end{align}

Applying a spinorial derivative to $\psi_1,...,\psi_{18}$ and projecting onto
the representation $(100) \times (20)$ (which we collectively denote by
$\D^Y$) we find 
\be
\begin{array}{ll}
\D^Y \psi_1 = -3i Y_6, &\D^Y \psi_2 = -3i Y_2, \\
\D^Y \psi_3 = 192 Y_1 -6i Y_7 + 3i Y_6, &\D^Y \psi_4 = 192 Y_1 +6i Y_3 -3i
Y_2, \\ 
\D^Y \psi_5 = 0, &\D^Y \psi_6 = 6i Y_8, \\
\D^Y \psi_7 = -16Y_1 + \frac{i}{2} Y_3, &\D^Y \psi_8 = -8 Y_1 + \frac{i}{12}
Y_2 + \frac{i}{12}Y_3 - \frac{i}{6}Y_4 + \frac{i}{6} Y_5, \\
\D^Y \psi_9 = -\frac{i}{2} Y_4, &\D^Y \psi_{10} = \frac{i}{6} Y_5 +
\frac{i}{12} Y_2 - \frac{i}{6} Y_3 + \frac{i}{12} Y_4, \\
\D^Y \psi_{11} = 16 Y_1 + \frac{i}{2} Y_8, &\D^Y \psi_{12} = -8Y_1 -
\frac{i}{4} Y_8, \\ 
\D^Y \psi_{13} = \frac{i}{2} Y_7, &\D^Y \psi_{14} = -12 Y_6 + 12 Y_7, \\
\D^Y \psi_{15} = -4 Y_2 + 4 Y_4 - 8 Y_5 + 16 Y_8, &\D^Y \psi_{16} = 4Y_2 -
4Y_3 +8 Y_5 - 8Y_8, \\ 
\D^Y \psi_{17} = 12 Y_2 - 12 Y_4, &\D^Y \psi_{18} = -24 Y_5 + 12 Y_2 - 12
Y_3. 
\end{array}
\ee
One can easily check that these relations are consistent with each other in
that $\D^Y \D$ must be identically zero.
The set $\{ \psi_3 ,\psi_4 ,\psi_5  ,\psi_8 ,\psi_{10} ,\psi_{11} ,\psi_{16}
,\psi_{17} \}$ can consistently be removed by field redefinition. Of the ten 
remaining $\psi$ the following linear combinations satisfy closure in the
$Y$-sense,
\begin{align}
&4i\psi_1 + 24i \psi_{13} + \psi_{14}, \\
&8i\psi_2 + 6i \psi_6 + 24i \psi_7 -24i \psi_9 - 48i \psi_{12} + 3\psi_{15} -
  \psi_{18}. 
\end{align} 
To check closure in the $Z$-sense we apply a spinorial derivative to these
$\psi$ followed by a projection onto the representation $(002) \times (12)$
(which we collectively denote by $\D^Z$). We find
\be
\D^Z \psi_1 = iZ_7, \hspace{30pt} \D^Z \psi_{13} = -\frac{i}{6}Z_7 +iZ_6,
\hspace{30pt} \D^Z \psi_{14} = 24 Z_6,
\ee
and so the first of the two linear combinations satisfies closure fully and is
hence a non-trivial element of the cohomology $H$. The second combination
fails $Z$-closure as one can see by the fact that the $\psi_6$ term is the
only term which makes a non-vanishing contribution to the constraint $Z_1$.

We found the program LiE \cite{lie} useful to check that we have the correct
number of terms in each representation.


\begin{thebibliography}{99}

\bibitem{hst83}
P. S. Howe, G. Sierra, P. K. Townsend,
{\sl Supersymmetry in Six Dimensions},
Nucl. Phys. B221 (1983) 331-348.

\bibitem{hs96}
P. S. Howe, E. Sezgin,
{\sl D=11, p=5},
Phys. Lett. B394 (1997) 62-66,
hep-th/9611008.

\bibitem{hsw97}
P. S. Howe, E. Sezgin, P. C. West,
{\sl Covariant Field Equations of the M Theory Five-Brane},
Phys.Lett. B399 (1997) 49-59,
hep-th/9702008.

\bibitem{cvgnw96}
M. Cederwall, A. von Gussich, B.E.W. Nilsson, A. Westerberg,
{\sl The Dirichlet Super-Three-Brane in Ten-Dimensional Type IIB Supergravity},
Nucl.Phys. B490 (1997) 163-178,
hep-th/9610148.

\bibitem{aps96}
Mina Aganagic, Costin Popescu, John H. Schwarz,
{\sl D-Brane Actions with Local Kappa Symmetry},
Phys.Lett.B393:311-315,1997,
hep-th/9610249.

\bibitem{cvgnsw96}
M. Cederwall, A. von Gussich, B.E.W. Nilsson, P. Sundell, A. Westerberg,
{\sl The Dirichlet Super-p-Branes in Ten-Dimensional Type IIA and Type IIB
  Supergravity},
Nucl. Phys. B490 (1997) 179-201,
hep-th/9611159.

\bibitem{blnpst97}
I. Bandos, K. Lechner, A. Nurmagambetov, P. Pasti, D. Sorokin, M. Tonin,
{\sl Covariant Action for the Super-Five-Brane of M-Theory},
Phys. Rev. Lett. 78 (1997) 4332-4334,
hep-th/9701149.

\bibitem{hklt03}
P. S. Howe, S. F. Kerstan, U. Lindstr\"om, D. Tsimpis,
{\sl The deformed M2-brane},
JHEP 0309 (2003) 013,
hep-th/0307072.

\bibitem{dk04}
J. M. Drummond, S. F. Kerstan,
{\sl Kappa-symmetric derivative corrections to D-brane dynamics},
JHEP10(2004)006,
hep-th/0407145.

\bibitem{cnt01a}
M. Cederwall, B. E. W. Nilsson, D. Tsimpis,
{\sl The structure of maximally supersymmetric Yang-Mills theory:
  constraining higher-order corrections},
JHEP 0106 (2001) 034,
hep-th/0102009.

\bibitem{cnt01c}
M. Cederwall, B. E. W. Nilsson, D. Tsimpis,
{\sl Spinorial Cohomology and Maximally Supersymmetric Gauge Theories},
JHEP 0202 (2002) 009,
hep-th/0110069.

\bibitem{sorokinreview}
D. P. Sorokin,
{\sl Superbranes and Superembeddings}
Phys.Rept.329:1-101,2000,
hep-th/9906142.

\bibitem{hs96a}
P. S. Howe, E. Sezgin,
{\sl Superbranes},
Phys. Lett. B390 (1997) 133-142,
hep-th/9607227.

\bibitem{kerstan02}
S. F. Kerstan,
{\sl Supersymmetric Born-Infeld from the D9-brane},
Class.Quant.Grav. 19 (2002) 4525-4536,
hep-th/0204225. 

\bibitem{at88}
O.D. Andreev, A.A. Tseytlin, 
{\sl Partition Function Representation for the Open Superstring
  Effective Action: Cancellation of Mobius Infinities and Derivatives
  Corrections to Born-Infeld Lagrangian}, 
Nucl.Phys.B311:205,1988.

\bibitem{bbg99}
C. P. Bachas, P. Bain, M. B. Green,
{\sl Curvatures Terms in D-Brane Actions and their M Theory Origin},
JHEP 9905:011,1999,
hep-th/9903210.

\bibitem{wyllard00}
N. Wyllard,
{\sl Derivative corrections to D-brane actions with constant
  background fields}, 
Nucl.Phys. B598 (2001) 247-275,
hep-th/0008125.

\bibitem{cdre02}
A. Collinucci, M. de Roo, M. G. C. Eenink,
{\sl Derivative corrections in 10-dimensional super-Maxwell theory},
JHEP 0301 (2003) 039,
hep-th/0212012.

\bibitem{dhhk03}
J. M. Drummond, P. J. Heslop, P. S. Howe, S. F. Kerstan,
{\sl Integral invariants in N=4 SYM and the effective action for
  coincident D-branes},
JHEP 0308 (2003) 016,
hep-th/0305202.

\bibitem{dre03}
M. de Roo, M. G. C. Eenink,
{\sl The effective action for the 4-point functions in abelian open
  superstring theory},
JHEP 0308 (2003) 036,
hep-th/0307211.

\bibitem{clr01}
S. Corley, D. A. Lowe, S. Ramgoolam,
{\sl Einstein-Hilbert action on the brane for the bulk graviton},
JHEP 0107 (2001) 030,
hep-th/0106067.

\bibitem{fotopoulos01}
A. Fotopoulos,
{\sl On $(\a')^2$ corrections to the D-brane action for non-geodesic
  world-volume embeddings},
JHEP 0109 (2001) 005,
hep-th/0104146.

\bibitem{aagg02}
F. Ardalan, H. Arfaei, M. R. Garousi, A. Ghodsi,
{\sl Gravity on Non-commutative D-branes},
Int.J.Mod.Phys. A18 (2003) 1051-1066,
hep-th/0204117.

\bibitem{wijnholt03}
M. Wijnholt,
{\sl On Curvature-Squared Corrections for D-brane Actions},
hep-th/0301029.

\bibitem{epple04}
F. Epple,
{\sl Induced Gravity on Intersecting Branes},
JHEP 0409 (2004) 021,
hep-th/0408105.

\bibitem{cms04}
B. Cabrera Palmer, D. Marolf, P. J. Silva,
{\sl Quantum Polarisation of D4-branes},
hep-th/0409034.



\bibitem{lie}
{\sl LiE},
A. M. Cohen, M. van Leeuwen, B. Lisser, LiE v 2.2.2,
http://young.sp2mi.univ-poitiers.fr/\symbol{126}marc/LiE/




\end{thebibliography}
\end{document}